\documentclass[doublecol]{epl2}

\title{Long memory constitutes a unified mesoscopic mechanism consistent with nonextensive statistical mechanics}
\shorttitle{Long memory constitutes a unified mesoscopic mechanism for nonextensivity} 

\author{Ananias M. Mariz\inst{1} \and Constantino Tsallis\inst{2,3}}

\institute{
  \inst{1} Departamento de Fisica Teorica e Experimental, Universidade Federal do Rio Grande do Norte, Natal-RN, Brazil\\
  \inst{2} Centro Brasileiro de Pesquisas Fisicas and \\National Institute of Science and Technology for Complex Systems, Rua Xavier Sigaud 150, 22290-180 Rio de Janeiro-RJ, Brazil\\
  \inst{3} Santa Fe Institute,
           1399 Hyde Park Road, Santa Fe, NM 87501, USA \\}

\pacs{05.45.-a}{Nonlinear dynamics}
\pacs{05.10.Gg}{Fokker-Planck equation,statistical physics}
\pacs{02.50.Ga}{Markov processes}

\abstract{
We unify two paradigmatic mesoscopic mechanisms for the emergence of nonextensive statistics, namely the multiplicative noise mechanism leading to a {\it linear} Fokker-Planck (FP) equation with {\it inhomogenous} diffusion coefficient, and the non-Markovian process leading to the {\it nonlinear} FP equation with {\it homogeneous} diffusion coefficient. More precisely, we consider the equation $\frac{\partial p(x,t)}{\partial t}=-\frac{\partial}{\partial x}[F(x) p(x,t)] + \frac{1}{2}D \frac{\partial^2}{\partial x^2} [\phi(x,p)p(x,t)]$, where $D \in {\cal R}$ and $F(x)=-\partial V(x) /\partial x$,  $V(x)$ being the potential under which diffusion occurs. Our aim is to find
 whether $\phi(x,p)$ exists such that the inhomogeneous linear and the homogeneous nonlinear FP equations become unified in such a way that the (ubiquitously observed) $q$-exponentials remain as stationary solutions. It turns out that such solutions indeed exist for a wide class of systems, namely when $\phi(x,p)=[A+BV(x)]^\theta [p(x,t)]^{\eta}$,
 where $A$, $B$, $\theta$ and $\eta$ are (real) constants.
 Our main result can be sumarized as follows: For $\theta \neq 1$ and arbitrary  confining potential
 $V(x)$, $p(x,\infty) \propto \left\lbrace 1-\beta(1-q)V(x)\right\rbrace ^{1/(1-q)} \equiv e_
 q^{-\beta V(x)}$,   where $q= 1+ \eta/(\theta-1)$.
The present approach unifies into a single  mechanism, essentially {\it long memory}, results currently discussed and applied in the literature.}

\begin{document}

\maketitle
One of the cornerstones of statistical mechanics is the functional connection
of the thermodynamic entropy with the set of probabilities $\{p_i\}$ of microscopic
configurations. For the celebrated Boltzmann-Gibbs (BG) theory, this central functional is given by
$S_{BG}=-k \sum_{i=1}^W p_i \ln p_i$,
where $W$ is the total number of microscopic states which are compatible
with the information that we have about the system. This powerful connection
is in principle applicable to a vast class of relevant systems, including (classical)
dynamical ones whose maximal Lyapunov exponent is positive, thus generically
warranting strong chaos, hence mixing in phase space, hence ergodicity
(in some sense, Boltzmann embodied all these features in his insightful
{\it molecular chaos hypothesis}).  Within this theory, it ubiquitously emerges the
Gaussian distribution
$ p_G \propto e^{-\beta x^2}  \quad (\beta >0) $.
Indeed, this important probabilistic form (i) maximizes the (continuous version of the)
entropy $S_{BG}=-k \int dx \, p(x) \ln [p(x)]$ under the basic constraints of normalizability
and finite width; (ii) constitutes the exact solution, for all values of space and time, of
the simplest form of the (linear and homogeneous) Fokker-Planck equation, in turn based on the simplest
form of the Langevin equation (which includes {\it additive noise}); (iii) is the $N \to\infty$
attractor of the (appropriately centered and scaled) sum of $N$ {\it independent} (or
weakly correlated in an appropriate sense) discrete or continuous random variables whose
second moment is finite (Central Limit Theorem, CLT); (iv) is the velocity distribution
(Maxwell distribution) of any classical many-body Hamiltonian system whose canonical
(thermal equilibrium with a thermostat) partition function is finite, i.e., if the interactions
between its elements are sufficiently short-ranged, or inexistent. The simplest probabilistic
model which realizes these paradigmatic features is a set of $N$ independent equal
binary random variables (each of them taking say the values 0 and 1 with probability 1/2).
The probability of having, for fixed $N$, $n$ $1$'s is given by
$\frac{N!}{n! \,(N-n)!}\, 2^{-N}$. Its limiting distribution is, after centering and scaling,
a Gaussian (as first proved by de Moivre and Laplace),
and its (extensive) entropy is the $BG$ one, since $S_{BG}(N)=N k\ln 2$.

What happens with the above properties when the correlations between say the
elements of a probabilistic model are strong enough (in the sense that they
spread over all elements of the system)? There is in principle no reason for
expecting the relevant limiting distribution to be a Gaussian, and the entropy
which is extensive (i.e.,  $S(N) \propto N$ for $N \gg 1$) to be $S_{BG}$. The purpose of
the present paper is to focus on such and related questions for a class of systems which are
ubiquitous in natural, artificial and even social systems, namely those which are
{\it scale-invariant} in a probabilistic sense which we shall define below.
Let us now discuss the frequent emergence of $q$-exponentials, defined as
\begin{equation}
P_q(x) = N_q[1-(1-q)\beta V(x)]^{1/(1-q)} = N_q \, e_q^{- \beta V(x)}\quad ,
\end{equation}
where $N_q$ is a normalization factor; $P_1(x)=N_1 \,e^{- \beta V(x)}$ is the standard BG case; for $V(x) \propto x^2$, $P_q(x)$ is a $q$-Gaussian,
which displays asymptotic power-laws and can be seen as a natural generalization of
the Gaussian ($q=1$).

At this point let us make a few remarks. (i) $q$-Gaussians appear as the exact solutions of paradigmatic non-Markovian
Langevin processes and their associated Fokker-Planck equations . Langevin equations
with both additive and multiplicative noise \cite{AnteneodoTsallis2003}, or Langevin equations
with long-range-memory \cite{FuentesCaceres2008}, lead respectively to
inhomogeneous linear \cite{Borland1998},   or  homogeneous nonlinear
\cite{PlastinoPlastino1995,TsallisBukman1996} Fokker-Planck equations (see also \cite{SchwammleCuradoNobre2007,SchwammleNobreCurado2007,SantosTsallis2010,AndradeSilvaMoreiraNobreCurado2010,BorlandPenniniPlastinoPlastino1999}).
(ii) $q$-CLT attractors are $q$-Gaussians \cite{UmarovTsallisSteinberg2008}.
(iii) The extremization of the entropy $S_q$ with norm and finite width constraints
yields
$q$-Gaussians, where
$S_q$ is a generalization of BG entropy,
namely  \cite{Tsallis1988,Tsallis2009}
\begin{equation}
  S_q = k \frac{1-\int dx\, [p(x)]^q }{ q-1} \;\;\;\; (q \in{R}; \, S_1=S_{BG})
\label{qentropy}
\end{equation}
This entropy is, for $q \ne 1$,  {\it nonadditive} (see \cite{Penrose1970} for the current definition of additivity), i.e., for arbitrary probabilistically independent systems $A$ and $B$, the equality $S(A+B)=S(A)+S(B)$ is not satisfied.
However, for many systems a value of $q$, denoted by $q_{\rm ent}$,
exists for which $S_{q_{\rm ent}}$ is {\it extensive}, i.e., $S_{q_{\rm ent}}(N) \propto N \;\;(N \gg 1)$.
As is well known, for all standard short-range-interacting many-body Hamiltonian systems,
we have $q_{\rm ent}=1$. However, some
systems
exist
for which $q_{\rm ent}<1$ \cite{Tsallis2004,TsallisGellMannSato2005,CarusoTsallis2008,Sarandy2009}.
(iv) Numerical indications \cite{PluchinoRapisardaTsallis2007} for the distributions of velocities in quasistationary states of long-range Hamiltonians \cite{AnteneodoTsallis1998} suggest $q$-Gaussians.
Further, experimental and observational evidence for $q$-Gaussians exists for the
motion of biological cells \cite{UpadhyayaRieuGlazierSawada2001,cellmove},
defect turbulence \cite{DanielsBeckBodenschatz2004}, solar wind
\cite{BurlagaVinas2005,BurlagaVinasAcuna2006}, cold atoms in dissipative
optical lattices  \cite{DouglasBergaminiRenzoni2006}, dusty plasma \cite{LiuGoree2008},
among others (see also \cite{NobreMonteiroTsallis2011}). Numerical indications are also available at the edge of chaos of unimodal
maps \cite{TirnakliBeckTsallis2007}.

A domain where the nonadditive entropy $S_q$ can be naturally incorporated is for describing anomalous diffusion like-phenomena. From modified Langevin equations, inhomogeneous linear or  homogeneous nonlinear Fokker-Planck equations have been  derived and used in order to obtain the  mesoscopic  dynamic evolution of  systems where such diffusion occurs.   

Let us consider here the nonlinear Fokker-Planck (FP) equation given by

\begin{equation}
\frac{\partial p(x,t)}{\partial t}=-\frac{\partial}{\partial x}[F(x) p(x,t)] + \frac{1}{2}D \frac{\partial^2}{\partial x^2} [\phi(x,p)p(x,t)] \,\
\end{equation}

\noindent where $D \in {\cal R}$ is the coeficient of diffusion, $F =  -\frac{\partial V(x)}{\partial x}$ the drift term, and $V$ a confining potential. We shall further assume the following wide connection:

\begin{equation}
\phi(x,p)=[g(V)]^\theta [p(x,t)]^{\eta} \,,
\end{equation}

\noindent with $g(V) = [A+BV]$,  $A$, $B$, $\theta$ and $\eta$ being real constants.

It should be noted that the present  equation  encompasses also  the general inhomogeneous  $(\theta \ne 0)$ nonlinear $(\eta \ne 0)$ case.
Our aim is to find whether a $q$-exponential $P_q(x)$ exists as a stationary solution (i.e., $ \lim_{t\to\infty}p(x,t)=P_q(x)$) of this FP equation such that the {\it inhomogeneous linear} and the {\it homogeneous nonlinear} FP equations become unified. 

The condition
$\frac{\partial p(x,t)}{\partial t}  = 0$ implies
\begin{equation}
\frac{\partial F(x)p(x,\infty)}{\partial x}  =  \frac{D}{2}\frac{\partial^2}{\partial x^2} [\phi(x,p)p(x,\infty)] \,.
\end{equation}

\noindent By assuming appropriate boundary conditions (basically $p(\pm \infty,\infty)=0$), changing variables and developing, we obtain
\begin{equation}
\frac{\partial \left[ g(V)^\theta p(V) ^{1+\eta}\right] }{\partial V} =  - \frac{2p(V)}{D} \,,
\end{equation}
where $p(V) \equiv p(V(x),\infty)$. It follows that
\begin{equation}
\frac{\partial p(V)}{\partial V}=-\frac{ 2g(V)^{-\theta} p(V)^{1-\eta} + \theta B D g(V)^{-1}p(V)}
{D(1+\eta)}\,.
\end{equation}

\noindent Now,  we will first consider the  $\theta \neq 1$ case.
Let us propose a solution of Eq.(7)  satisfying

\begin{equation}
g(V)^{1-\theta} p(V)^{-\eta}  = C
\label{eq2}
\end{equation}

\noindent with  $\partial C/\partial V = 0$.  By direct substitution in Eq. (7), we can easily show that $p(V) = P_q(V)$, as  given by Eq. (1), with

\begin{equation}
q = 1+ \frac{\eta}{\theta - 1}\,,
\label{eq3}
\end{equation}

\noindent and
\begin{equation}
\beta = \frac{2C+\theta BD}{AD(1+\eta)}\,,
\end{equation}

\noindent where
\begin{equation}
C= \left[ AN_q^{q-1}\right] ^{1-\theta} \,.
\end{equation}

These results reproduce, for $\theta =0$, the homogeneous nonlinear case discussed in \cite{PlastinoPlastino1995,TsallisBukman1996, FuentesCaceres2008}, i.e.,  $q=1- \eta$ and $N_q^{1-q} \beta = 2/ D(1+\eta)$,
which leads to normal diffusion for $\eta =0$.

Eq. (9) shows that, for $\eta \neq 0$, the limit $\theta \rightarrow 1$ corresponds to a singularity in the $ |q| $ value. However, in this limit, we still find $q$-exponentials as stationary solutions of the above FP equation, in two different situations:

a) For $\eta$ considered as a function of $\theta$ with the leading term  given by $\eta \sim \alpha(\theta-1)^\delta$, the analytical extension of the solution given by Eqs. (8-11) results in $q$-exponentials presenting the indices  $q=1$ for $\delta>1$ and $q=1+\alpha$ for $\delta=1$, with  $\beta=\left( 2 +  BD\right)/AD$ in both cases.

b) An isolated solution (which is not a particular case of the previous ones) can also be found by setting $\eta=0$ and $\theta = 1$ in Eq. (7).
By so doing we obtain

\begin{equation}
\frac{\partial p(V)}{\partial V}=- \frac{\left[ 2 + BD\right] g(V)^{-1} p(V)}{D} \,.
\end{equation}
Now, imposing the relation

\begin{equation}
g(V)p(V)^{q-1}=\bar C \,.
\end{equation}
where $\bar C$ is independent of the potential $V$. Following the same procedure used to obtain Eqs. (9) and (10), we finally have $p(V)=P_q(V)$, with
\begin{equation}
q=\frac{\left( 1+BD\right)}{\left( 1+BD/2\right)} \,,
\end{equation}

and

\begin{equation}
\beta = \frac{\left( 2+BD\right)}{AD} \,.
\end{equation}

It should be noticed that this case (($\eta,\theta)=(0,1)$) recovers  previous results already obtained in \cite{Borland1998,AnteneodoTsallis2003}.

As final remarks, we note that:

(i) For  $A \neq 0$, we may take $A=1$ without loss of generality, if the $(B,D)$ parameters are properly rescaled;

(ii) It is known \cite{Tsallis2009}, that the $q$-Gaussians, which emerge for $V(x) \propto x^2$ are normalized only for $q<3$, which implies the following restrictions to the values of the parameters of the FP equation: (a) For $\theta \neq 1$, $\frac{\eta}{\theta - 1}\,< 2$, (b) For $\theta \rightarrow 1$, $\alpha<2$ if $\delta=1$ and $BD>-2$ or $BD<-4$. For each  V(x), an analysis of integrability of $P_q$ must be performed, in order to establish the accepted range of values of the parameters ($\eta,\theta,B,D,q$) (e.g., if $V(x) \propto |x|^\rho$, in the limit $|x| \to\infty$, then it must be $q<1+\rho$);

(iii) We stress that the  Eq. (3) used in this paper corresponds to the It${\hat{o}}$ form of a generalized FP equation (a Stratonovich approach has already been examined \cite{BorlandPenniniPlastinoPlastino1999}, and the existence of $q$-exponential distributions as stationary solutions has also been proved).

Let us emphasize that Eq. (9), here presented for the first time, enables us to analyze in an unique way the present general inhomogeneous nonlinear case, which contains, as particular cases, several FP equations used to describe a large class of nongaussian natural systems. 
This unification constitutes the main result of this paper.

To summarize, the {\it nonlinear inhomogeneous} FP process given by Eq. (3) has, as a stationary solution for any confining potential $V(x)$, a probability distribution given by the $q$-exponential $P_q(V)$, with finite values of $q$ and $\beta$ (see Fig. 1 and Table 1). These results exhibit that we may retain long memory (basically, the probability distribution longstandingly maintains memory of its form at $t=0$) as a unified mesoscopic mechanism consistent with nonextensive statistical mechanics.

\begin{figure}
\begin{center}
\includegraphics[width=8.5cm]{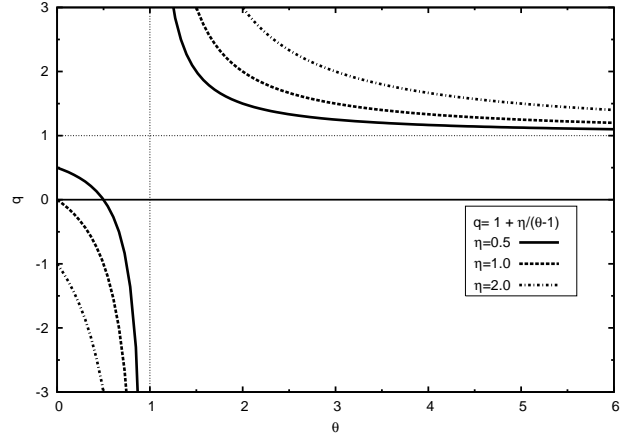}
\end{center}
\vspace{-0.5cm}
\caption{$q = 1+\frac{\eta}{\theta - 1}\,$ as a function of $\theta$ for selected values of $\eta$ (the upper bound $q=3$ herein shown corresponds to a potential $V(x) \propto x^2$, i.e., $\rho=2$). For the special point ($\theta \rightarrow 1$, $\eta 
\rightarrow 0$), there are three different solutions 
for   $q$, namely $q=1$, $q=1+\alpha $, and $q = (1+BD)/(1+BD/2)$ (see the text).}
\end{figure}

We acknowledge useful remarks by A.O. Sousa, as well as partial financial support by Fapern, Faperj and CNPq (Brazilian agencies).

\onecolumn

\begin{table}[htbp]
\hspace{3.5cm}
\begin{tabular}{c||c|c||}
 {\bf FOKKER}             &  {\bf Linear}                                     &    {\bf Nonlinear}   \\
 {\bf -PLANCK}                         &     $(\eta \to 0)$                                  & $(\eta \ne 0)$    \\
[2mm] \hline\hline
                                            &                                       &                                                               \\
{\bf Homogeneous}                               &  $q=1$                    & $q=1-\eta$    \\
$(\theta=0)$                       &                                       &                                             \\
[3mm] \hline
                                             &                                      &                                                               \\
{\bf Inhomogeneous}       &   If  $\eta \sim \alpha(\theta-1)^\delta$ and $\theta \rightarrow 1$, & $q=1+\frac{\eta}{\theta-1}$   \\
$(\theta \ne 0)$                 &   $q=1$  for  $\delta>1$                                         & \\
                                             &       $q=1+\alpha$  for  $\delta=1$                    & \\
                                             &                                                          &\\
                                            &    If $\eta$ is strictly $0$ and $\theta \rightarrow 1$,    & \\
                                            &    $q=\frac{2(1+BD)}{2+BD}$                                       & \\
[2mm] \hline \hline
\end{tabular}

\caption{Nonlinear inhomogeneous Fokker-Planck equation with $q$-exponential  stationary-state distribution. The present $\theta=0$ result recovers that of Eqs. (2.8) and (2.9)  of \cite{PlastinoPlastino1995} with the notation correspondence $(F,Q,D,K)$ in \cite{PlastinoPlastino1995} $\leftrightarrow (P,D,N_q,F)$ here. It also recovers that of \cite{TsallisBukman1996} by using there $\mu=1$ and $\nu=1+\eta$. The result presented by the isolated solution at $(\theta,\eta)=(1,0)$ recovers that  cited (for the It${\hat{o}}$ approach) in \cite{AnteneodoTsallis2003} with the notation correspondence $(\frac{M}{\tau})$  in\, \cite{AnteneodoTsallis2003} $\leftrightarrow (\frac{BD}{4})$ here. A comparison is also possible through Eq. (11) of \cite{Borland1998} by using the notation correspondence $(K,D,U(x))$ in \cite{Borland1998} $\leftrightarrow (F,Dg(V)^\theta,V(x))$  here. }
\label{table3.1}
\end{table}

\end{document}